\documentclass[a4paper,11pt]{article}
\pdfoutput=1 % if your are submitting a pdflatex (i.e. if you have
\usepackage{jinstpub} % for details on the use of the package, please
\newcommand{\neqcm}{\ensuremath{\mathrm{n}_{\mathrm{eq}}/\mathrm{cm}^2}}

\title{\boldmath Optimization of thin n-in-p planar pixel modules for the ATLAS upgrade at HL-LHC }

\author[a,1]{A. Macchiolo,\note{Corresponding author.}}
\author[a]{J. Beyer,}
\author[a]{A. La Rosa,}
\author[a]{R. Nisius}
\author[a]{and N. Savic}

\affiliation[a]{Max-Planck-Institut f\"ur Physik,\\F\"ohringer Ring 6, 80805 Munich, Germany}

\emailAdd{Anna.Macchiolo@mpp.mpg.de}

\abstract{The ATLAS experiment will undergo around the year 2025 a replacement of the tracker system in view of the high luminosity phase of the LHC (HL-LHC) with a new 5-layer pixel system.
Thin planar pixel sensors are promising candidates to instrument the innermost region of the new pixel system, thanks to the reduced contribution to the material budget and their high charge collection efficiency after irradiation. The sensors of  50-150 $\mu$m thickness, interconnected to FE-I4 read-out chips, have been characterized with radioactive sources and beam tests. In particular active edge sensors have been investigated. The performance of two different versions of edge designs are compared: the first  with a bias ring, and the second one where only a floating guard ring has been implemented.
The hit efficiency at the edge has also been studied  after irradiation at a fluence of $10^{15}$ \neqcm.
Highly segmented sensors will represent a challenge for the tracking in the forward region of the pixel system at HL-LHC. In order to reproduce the performance of 50x50 $\mu$m$^2$ pixels at high pseudo-rapidity values, FE-I4 compatible planar pixel sensors have been studied before and after irradiation in beam tests at high incidence angles with respect to the short pixel direction. Results on the hit efficiency in this configuration are discussed  for different sensor thicknesses.
}


\proceeding{8th International Workshop on Semiconductor Pixel Detectors for Particles and
Imaging.\\
  5-9 September 2016\\
  Sestri Levante}

\begin{document}
\maketitle
\flushbottom

\section{Introduction}
\label{sec:intro}
The ATLAS experiment at CERN is planning a complete replacement of the tracking detector around 2024-2026 in view of the High Luminosity phase of the LHC (HL-LHC). The istantaneous luminosity is foreseen to be increased to 5-7$\times$10$^{34}$ cm$^{-2}$s$^{-1}$ and after ten years of running the integrated luminosity is expected to be around 3000 fb$^{-1}$ \cite{BenSmart}. The new pixel system will be composed by 5 layers, with the two innermost ones 
foreseen to be replaced at half of the Inner Tracker (ITk) lifetime. The maximum fluence expected before replacement in the innermost layer is around 1-1.2$\times10^{16}$ \neqcm \, \cite{STRIP_TDR}.  A higher pixel granularity with respect to the 50x400 and 50x250 $\mu$m$^{2}$ cell size presently employed in the ATLAS pixel detector \cite{ATLASPixel} will be necessary to cope with the increased particle rate, and the new read-out chip, being developed by the RD53 Collaboration in 65 nm CMOS technology,  will have a pitch  of 50x50 $\mu$m$^{2}$ \cite{RD53}.

In this paper thin planar pixel sensors, based on the n-in-p technology, are investigated as possible candidates to instrument all the layers of the ITk pixel system, and their electrical characterization before and after irradiation is shown in a sensor thickness range between 50 and 150 $\mu$m.
The main benefits of the use of thin planar pixel sensors are their high production yield, reduced contribution to the material budget and higher radiation resistance thanks to the higher electric field that can be established in the bulk with respect to thicker sensors at the same applied bias voltage.
The radiation resistance of 100 $\mu$m thin planar sensors has been proven up to a fluence of 10$^{16}$ \neqcm  \, and a power dissipation of 25-50 mW/cm$^2$ has been estimated at an operational temperature of -25$^o$C \cite{IWORID}.

\section{Thin planar pixel sensor productions}
\label{sec:thinpix}

Different production methods for thin planar pixel sensors are being investigated at the moment, from the established use of Silicon on Insulator (SOI) wafers to a new technology employed at CiS \cite{CIS} that does not foresee the use of a handle wafer. For all the sensors here described the inter-pixel isolation is achieved by means of a homogeneous p-spray realized with a maskless low dose boron implantation.

\begin{figure}[htbp]
\centering % \begin{center}/\end{center} takes some additional vertical space
\includegraphics[width=.45\textwidth]{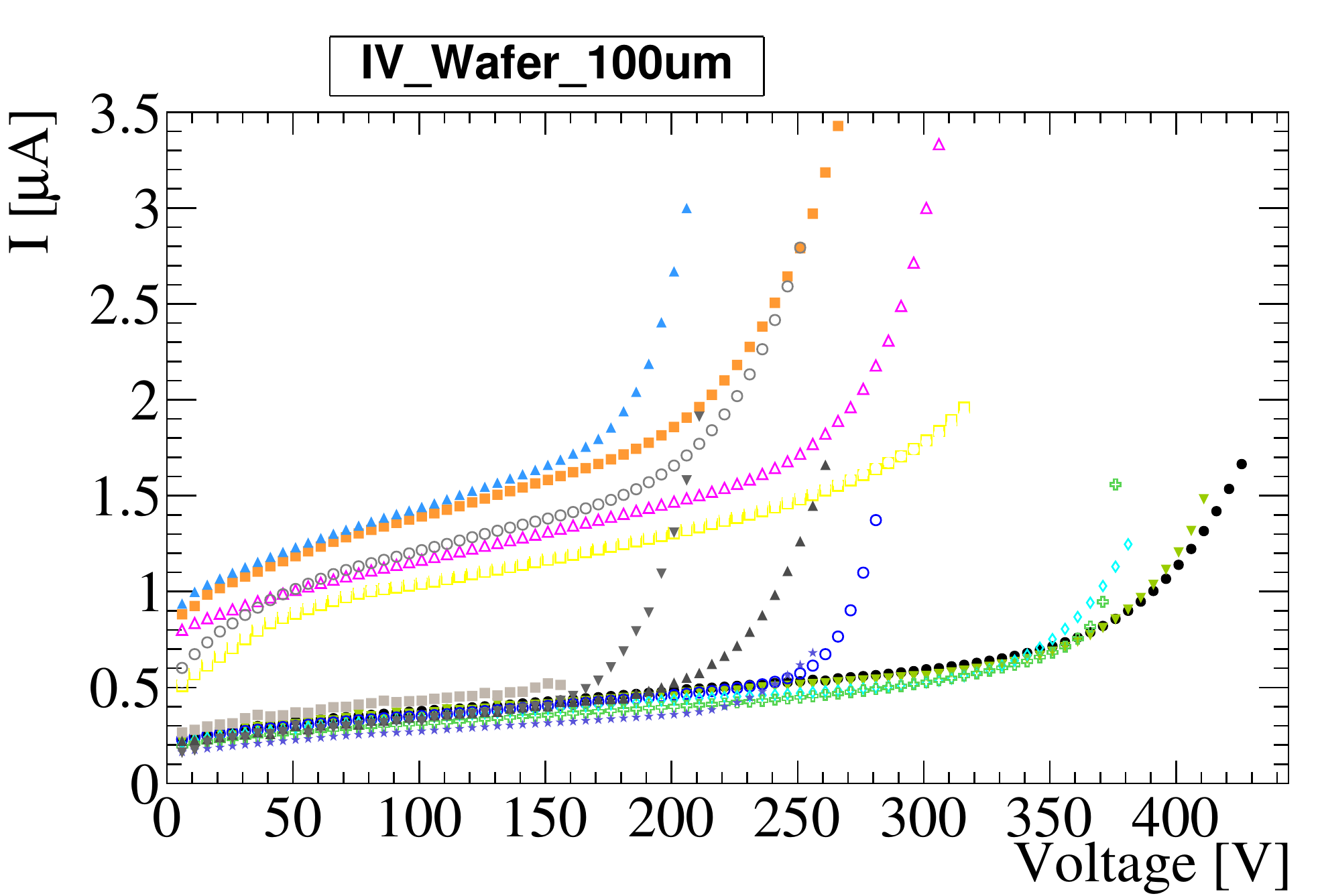}
\qquad
\includegraphics[width=.45\textwidth,origin=c]{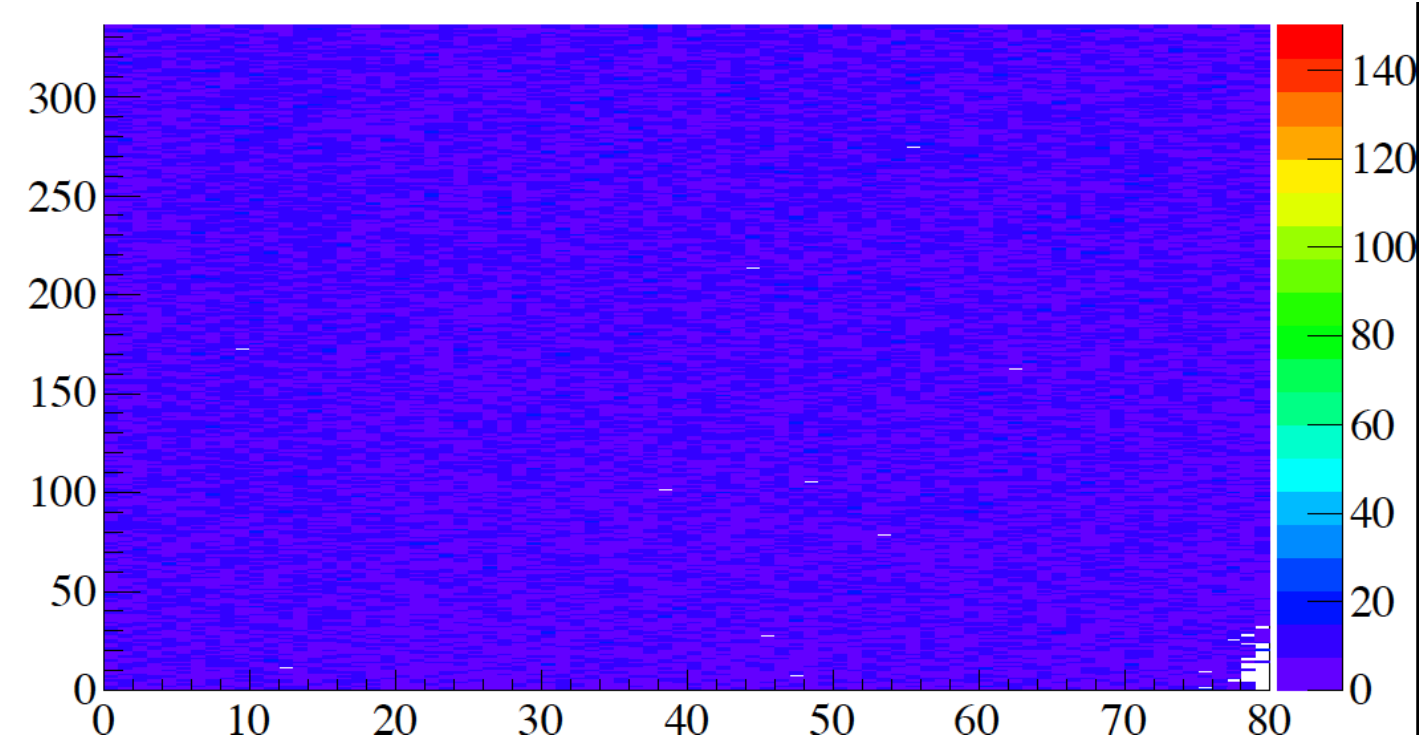}
% "\includegraphics" from the "graphicx" permits to crop (trim+clip)
% and rotate (angle) and image (and much more)
\caption{\label{fig:CisProduction} {Left side: IV curves measured wtith 100 $\mu$m thin bare sensors of the CiS production with backside cavities. The sensors are of two different sizes, for single FE-I4 chip or four chip assemblies. The depletion voltage for this thickness is around 10 V. Right side: hit map obtained with a Cadmium source scan on a FE-I4 module (336 rows $\times$ 80 columns). The map shows a good uniformity of the number of hits per pixel, except on the bottom right corner where the white color indicates that no hits were recorded.}}
\end{figure}

The thinning  technology at CiS is based on 
anisotropic potassium hydroxide (KOH) etching to create backside cavities in the wafer leaving thicker frames around each single structure.
The process, described in more detail in \cite{Vienna}  has the potential to be more cost-effective compared to standard techniques involving the use of  SOI wafers.  The first test has been performed on 4-inch wafers of 525 $\mu$m starting thickness, where the area below each sensor has been thinned to 100 or 150 $\mu$m. IV curves performed after dicing on 100 $\mu$m thick sensors are shown in Fig.\ref{fig:CisProduction}, with the breakdown voltages around ten times higher than the depletion voltage of 10 V for these devices. 
This first production has also been used to investigate two different Under Bump Metallization (UBM) technologies at CiS, involving thin film deposition of Nickel-Gold or Platinum layers as post-processing steps to allow for solder adhesion to the pixel Aluminum pads. These UBM methods are both compatible with the SnAg bump technology for the read-out chip wafers, developed at the Fraunhofer Institute IZM. 
Modules with the two UBM types, flip-chipped at IZM \cite{IZM}, have been investigated by means of scans with radioactive sources and beam tests at CERN-SPS.
The homogenous number of hits measured per channel with a Cd source scan, shown in the right plot of Fig.\ref{fig:CisProduction}, hints to a very high interconnection yield.
Given the good results obtained with this 4-inch wafer production, including mainly FE-I4 compatible sensors, a second one, on 6-inch wafers, has been started to realize sensors to be interconnected to the RD53A prototype chip, with pixel cell sizes of 50x50 $\mu$m$^{2}$ and 25x100 $\mu$m$^{2}$.

\begin{figure}[ht]
\centering
\includegraphics[width=0.8\columnwidth]{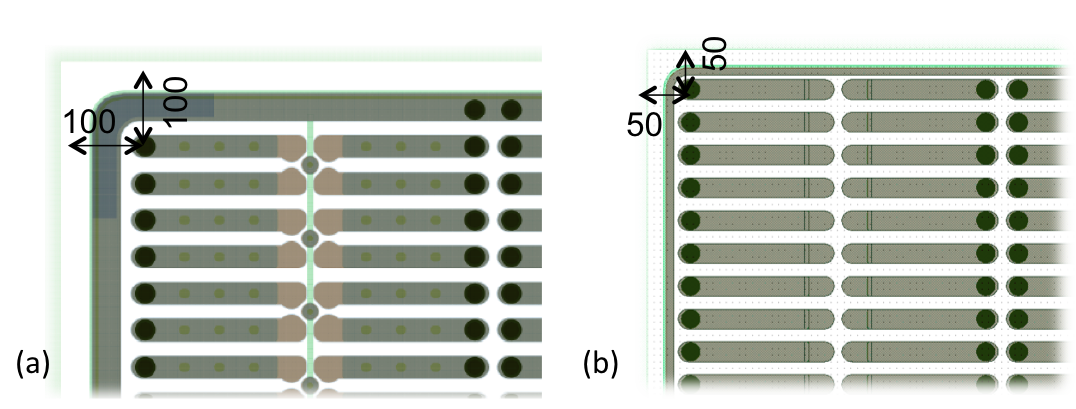}
\caption{Left: Slim edge design with 100 $\mu$m distance between the last pixel implant and the edge, it implements a bias ring and punch-through structures. Right: Active edge design with 50 $\mu$m distance between the last pixel implant and the edge, only a floating bias ring is included. }
\label{fig:ADVACAMEdges}
\end{figure}

An alternative method to produce thin sensors is the use of SOI wafers where the backside support is etched away at the
end of the processing of the front-side. With this technology
a production at ADVACAM \cite{ADVACAM} employing 6-inch Float Zone (FZ)
wafers was organized to investigate planar n-in-p pixel sensors with active edges. Three different sensor thicknesses have been 
chosen:  50,100 and 150 $\mu$m. Trenches are realized around the perimeter of each sensor with Deep Reactive Ion Etching (DRIE) using the mechanical support offered by the handle wafer of the SOI stack. A  boron implantation is performed  in the trenches with a four-quadrant ion implantation \cite{Kalliopuska}, allowing to extend the depleted volume up to the edges.
After the UBM processing, the handle wafer is removed, singularizing the structures. Also in this case, inter-pixel isolation is achieved by mean of a homogeneous p-spray.
A slim edge and an active edge design have been implemented differing by the distance  of the last pixel implant to the sensor edge, indicated as d in the following. The slim edge version  with d= 100 $\mu$m (Fig.\ref{fig:ADVACAMEdges} (a)), includes a bias ring and a punch-through biasing structure. 
The active edge design (Fig.\ref{fig:ADVACAMEdges} (b)) characterized by d= 50 $\mu$m includes only one floating guard ring (GR). No punch-through structure is implemented in this design. 

\begin{figure}[ht]
\centering
\includegraphics[width=\columnwidth]{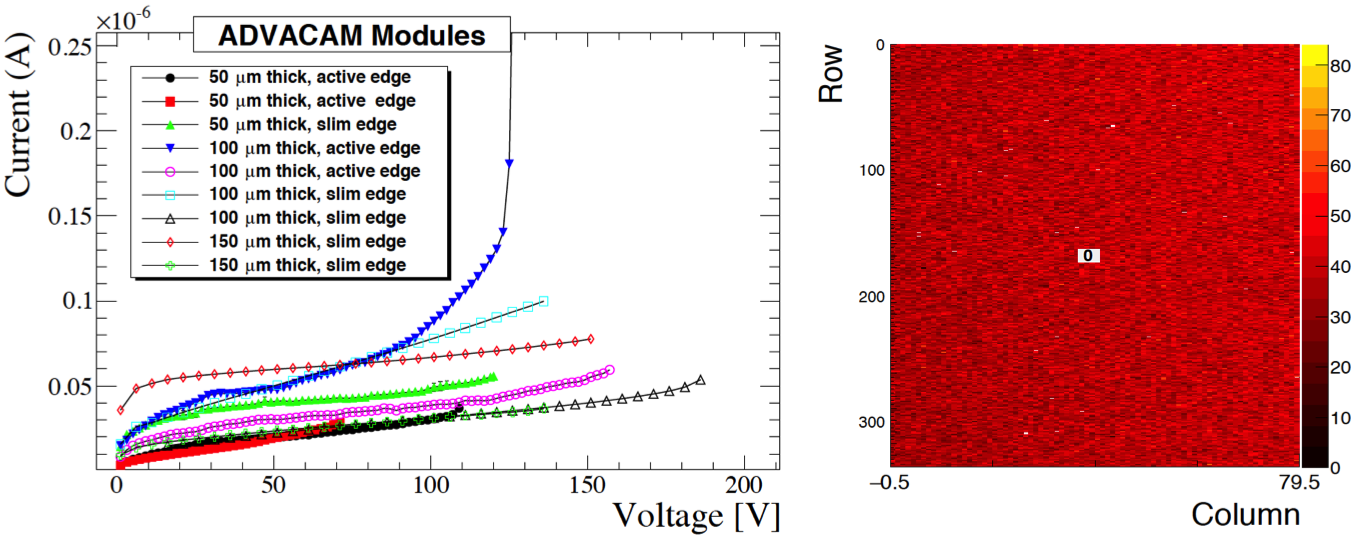}
\caption{Left: IV curves measured on FE-I4 modules  with ADVACAM sensors with a thickness of 50, 100 and 150 $\mu$m. Right: Hit map  obtained with a Cadmium source using a 50  $\mu$m thin sensor interconnected to a FE-I4 chip.}
\label{fig:ADVACAM_IV}
\end{figure}

Results of a first electrical characterization of all sensor types after flip-chipping to FE-I4 chips are summarized in Fig.\ref{fig:ADVACAM_IV} and show a good performance, in particular for the case of the 50 $\mu$m thin sensors, where the depletion voltage has been measured to be lower than 10 V. The IV curves on the left plot indicate typical breakdown voltages above 100 V, and very low levels of the leakage current, around 10 nA/cm$^2$. The hit map in the right plot, obtained by means of a Cadmium source scan with a 50 $\mu$m thick sensor, hints to a very low level of disconnected channels, given the homogenous hit counts per pixel. The measurements presented in this paper are based on devices with an electroplated CuAu UBM, that resulted in a better performance in terms of noise and interconnection yield with respect to thin film UBM that was also investigated with active edge sensors from the same wafers.

\section{Test beam characterization of active edge sensors}
\label{sec:active}

\begin{figure}[ht]
\centering
\includegraphics[width=0.8\columnwidth]{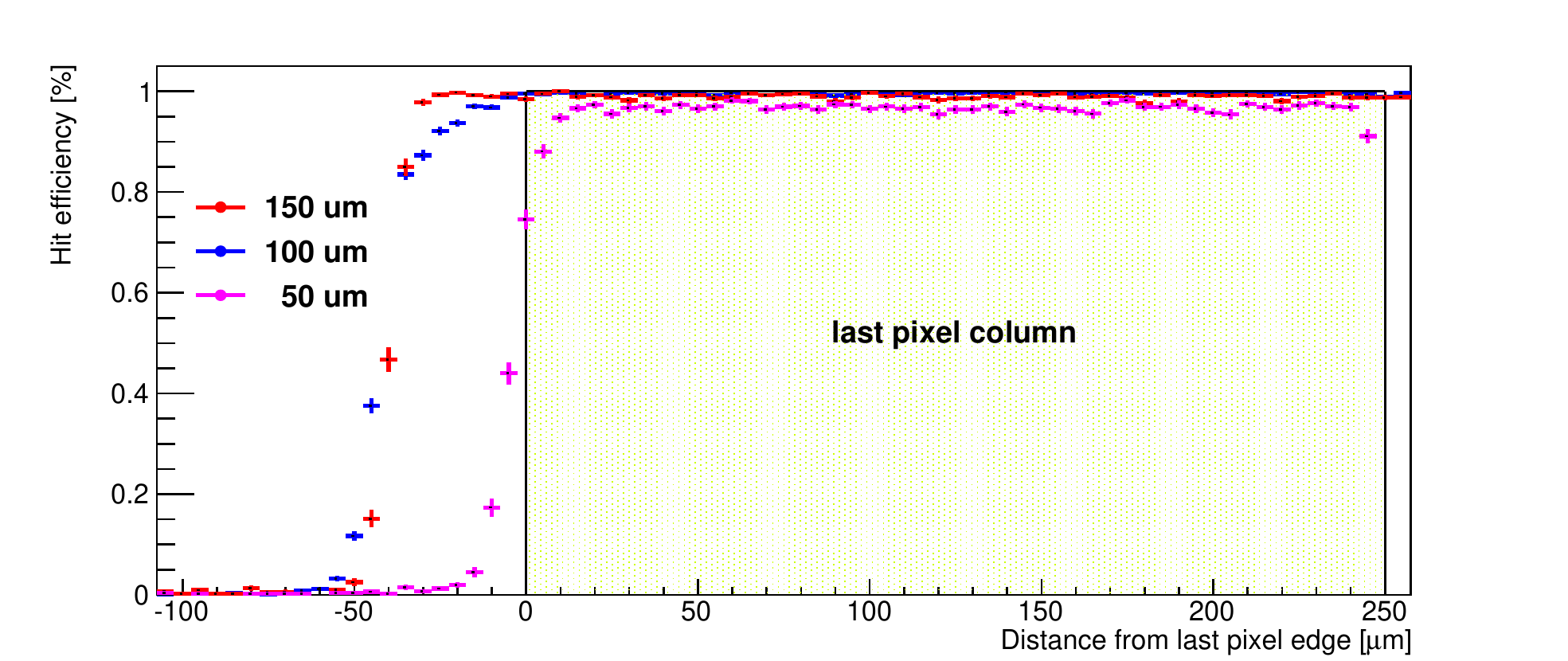}
\caption{Hit efficiency measured on not irradiated active edge sensors of 50, 100 and 150 $\mu$m thickness. The values are averaged among the pixels of the edge column. The yellow rectangle indicates the position of the pixel n$^+$ implant of the last pixel column terminating at x=0 while the sensor phisically ends at x=-50 $\mu$m.}
\label{fig:ActiveEdgeEff}
\end{figure}

The performance of the active edge sensors has  been investigated with beam tests  at CERN-SPS by using telescopes of the EUDET family \cite{EUDET}.
Fig.\ref{fig:ActiveEdgeEff} shows a comparison of the hit efficiency measured at a beam test at CERN-SPS with not-irradiated active edge sensors of different thickness. It can be observed that while the edge in the 100 and 150 $\mu$m thick sensors is almost fully efficient,  the 50 $\mu$m thick sensor is only efficient up to the end of the last pixel implant. Moreover also in this area the hit efficiency is 97\%, lower than for the thicker sensors, with the threshold tuned to a nominal value of 800 e$^-$. Further measurements are being carried out to determine the efficiency with an optmized tuning at lower values of the threshold. 

\begin{figure}[ht]
\centering
\includegraphics[width=0.7\columnwidth]{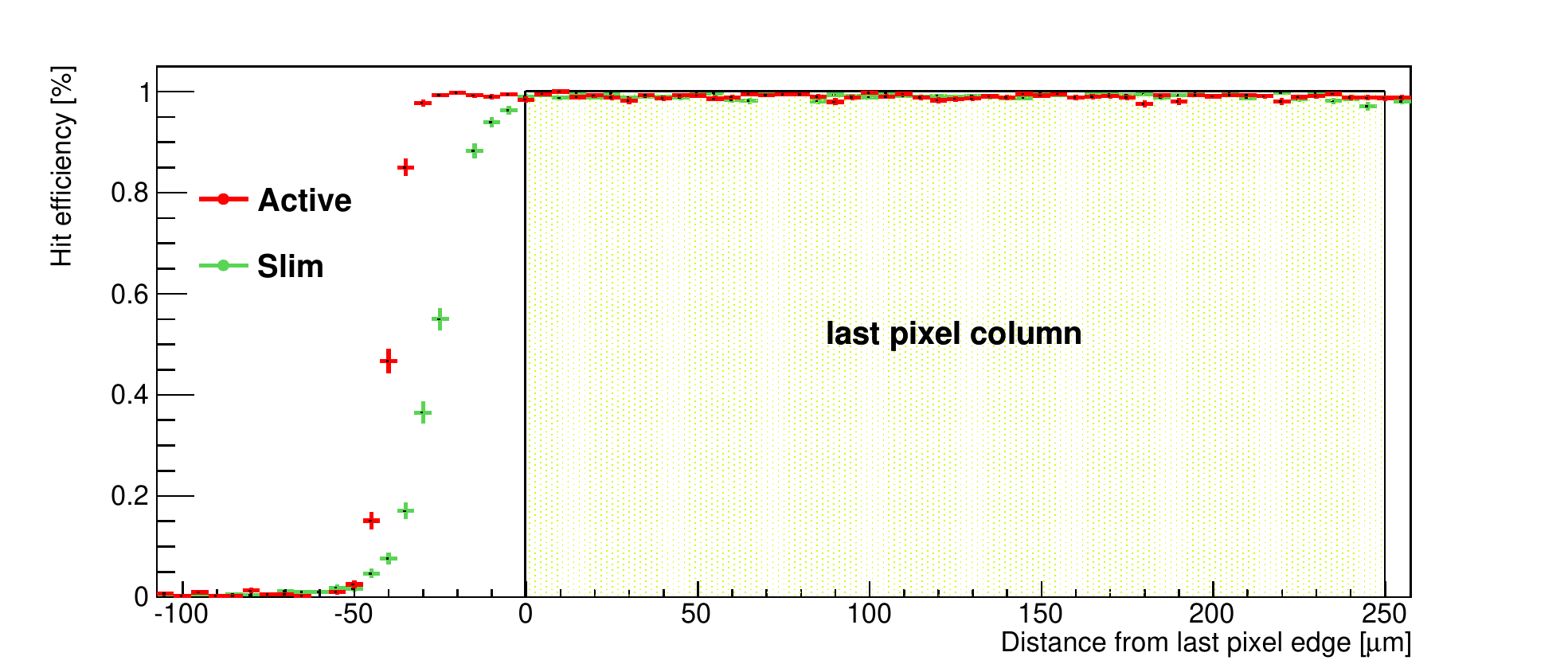}
\caption{Hit efficiency measured on not irradiated 150 $\mu$m thick sensors with active or slim edges. The values are averaged among the pixels of the edge column. The yellow rectangle indicates the position of the pixel n$^+$ implant of the last pixel column terminating at x=0 while the sensor phisically ends at x=-50 $\mu$m for the active edge design and at x=-100 $\mu$m for the slim edge design.}
\label{fig:Edge150Eff}
\end{figure}

The different performance of an active or slim design is shown in Fig.\ref{fig:Edge150Eff}  for  150 $\mu$m thick sensors, where it can be noticed that the slim edge design is only partially efficient in the edge region, given the fact that the charge collected by the bias ring, kept at ground potential by the read-out chip through two sets of bumps, is not amplified and further read-out.

\begin{figure}[ht]
\centering
\includegraphics[width=0.7\columnwidth]{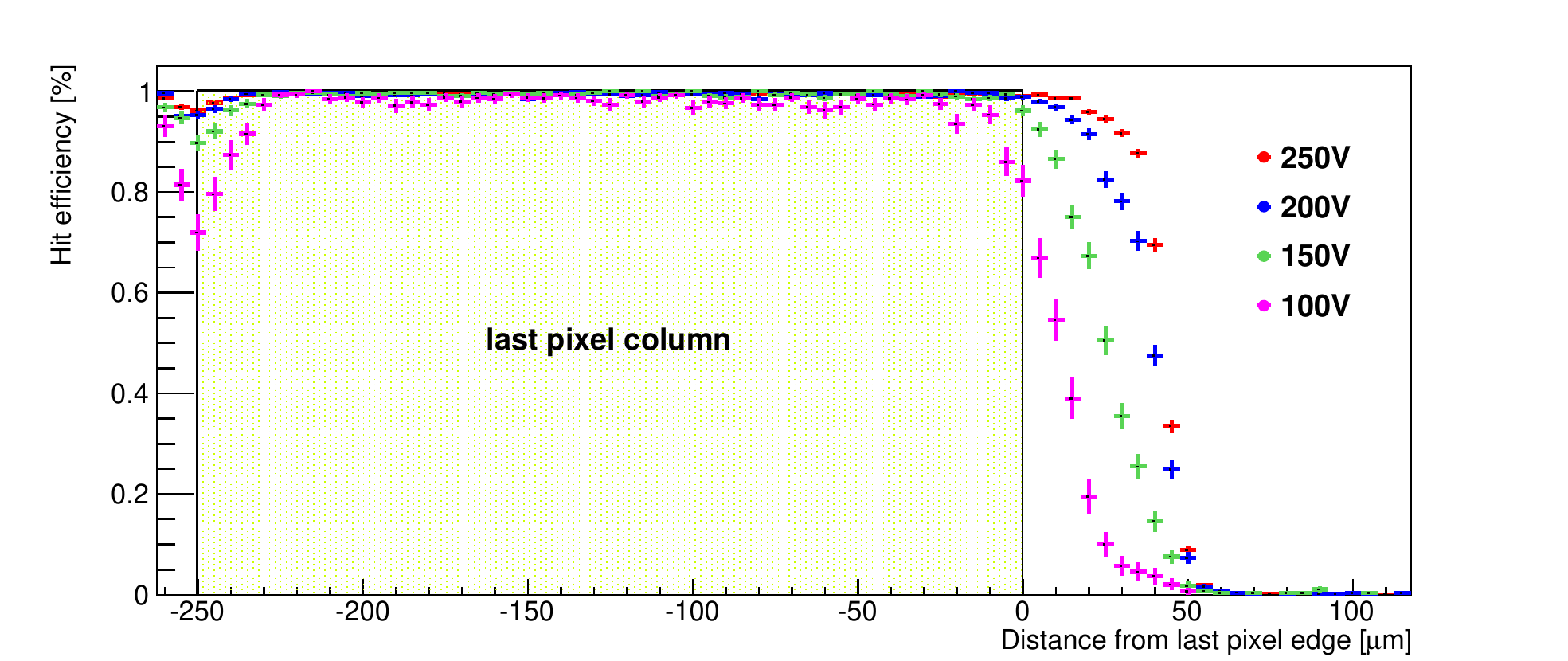}
\caption{Hit efficiency measured measured with a 150 $\mu$m thick sensor with active  edges after irradiation at a fluence of $10^{15}$ \neqcm  .}
\label{fig:ActiveEdgeIrr}
\end{figure}

\begin{figure}[ht]
\centering
\includegraphics[width=0.67\columnwidth]{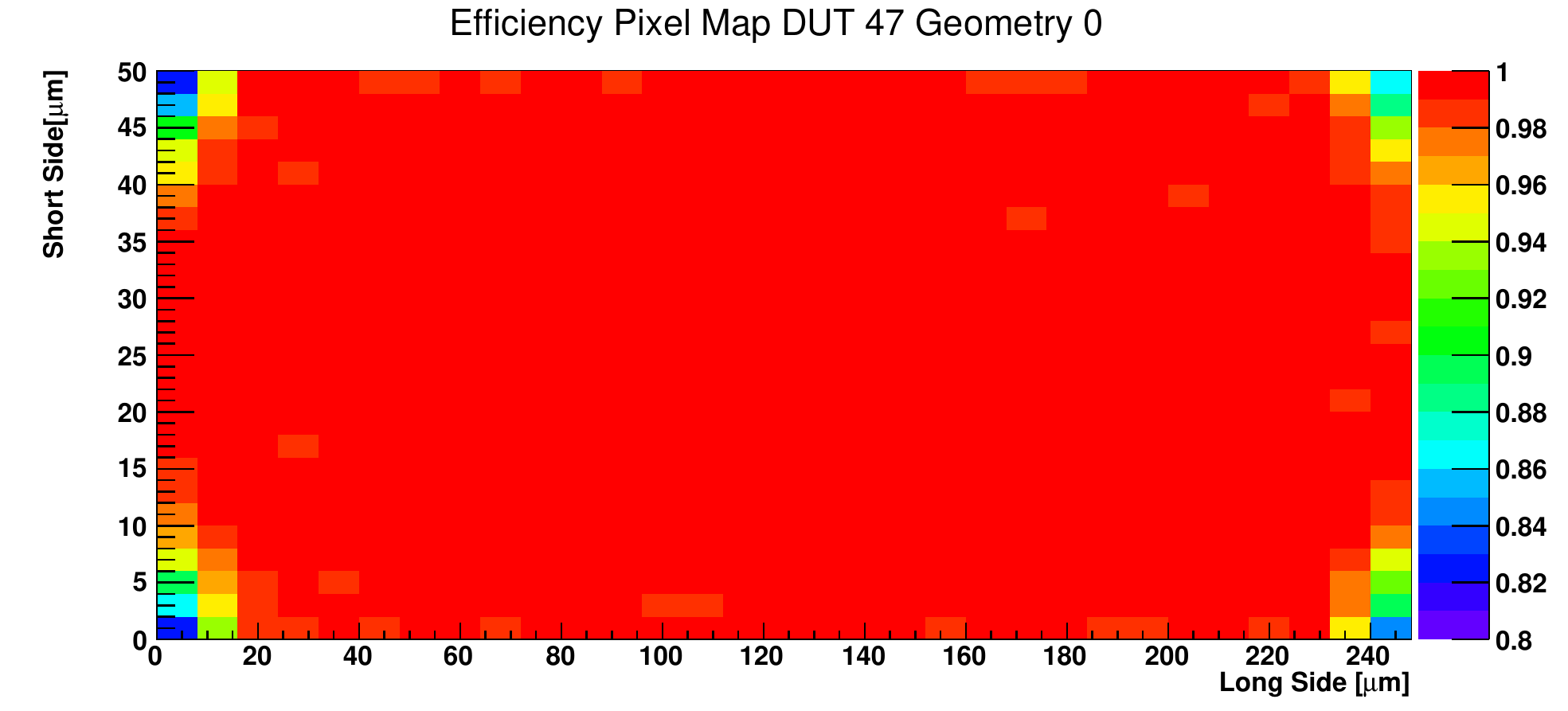}
\caption{In-pixel hit efficiency map of a  150 $\mu$m thick sensor with active edges at a bias voltage of 250 V, after irradiation at a fluence of $10^{15}$ \neqcm  .}
\label{fig:IrrPixelMap}
\end{figure}

The performance of active edge devices has also been studied after irradiation at a fluence of $10^{15}$ \neqcm \, with 24 MeV protons at the MC40 Proton Cyclotron in Birmingham \cite{UoB}, and the corresponding hit efficiency is shown in Fig.\ref{fig:ActiveEdgeIrr}. Increasing the bias voltage up to a value of 250 V, the edge efficiency is fully recovered and also the efficiency loss due to charge sharing between neighbouring pixels, visible in Fig.\ref{fig:IrrPixelMap} at the four corners of the pixel cell, is partially reduced. 

\section{Tracking performance at high incidence angle}
\label{sec:eta}
\begin{figure}[ht]
\centering
\includegraphics[width=0.55\columnwidth]{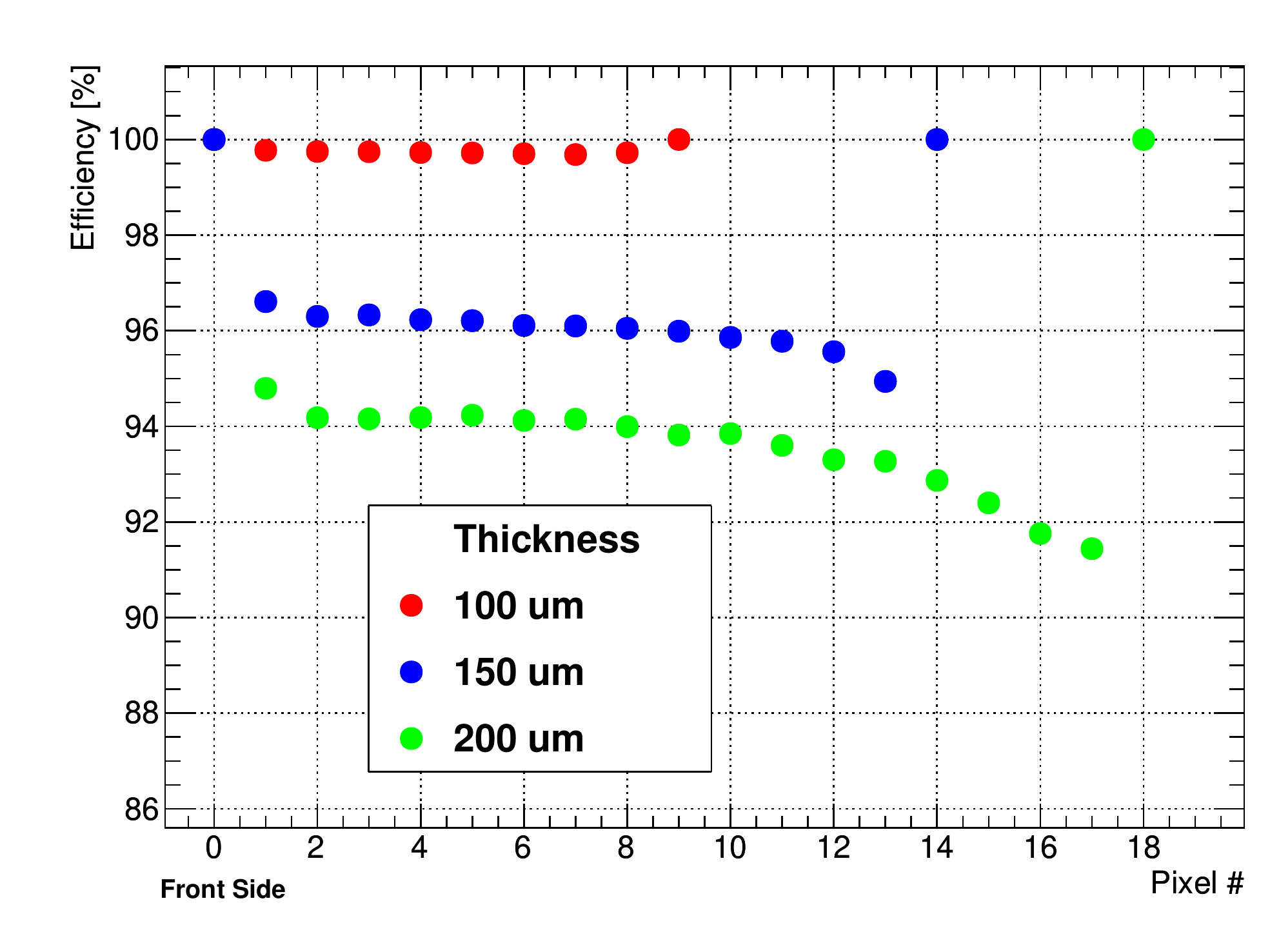}
\caption{Hit efficiency as a function of pixel number 
for not irradiated FE-I4 modules with a sensor thickness of 100, 150, 200  $\mu$m, inclined by 76-78$^o$  with respect to the beam
direction. The efficiencies of the first and last pixels are by construction 100\% since they define the start and
the end of the cluster.}
\label{fig:HighPhi}
\end{figure}

Smaller pixel cells are challenging for the tracking particles in regions of high pseudo-rapidity (high  $\eta$).
To make predictions on the performance of a 50x50 $\mu$m$^2$ pixel cell at high incidence angle, FE-I4 modules were
arranged such that the particles of the beams at DESY and CERN were crossing the pixel along the
short side (50 $\mu$m), at an inclination corresponding to  $\eta$ values around 2-2.5.  
The hit efficiency for each single pixel in the cluster is shown in Fig.\ref{fig:HighPhi}
for sensors of 100, 150 and 200 $\mu$m thickness. The single hit efficiency
is defined as the total number of hits divided by the cluster length not
including the entrance and exit pixels that are 100\% efficient by construction.
All modules were tuned to a threshold of 1000 e$^-$. 
It can be observed that the cluster size depends strongly on the sensor thickness and that the hit efficiency is higher for the thinner sensors,
possibly due to the reduced effect of charge sharing for lateral diffusion to the neighbouring pixels. 
\section{Conclusions}
Thin planar n-in-p pixel modules, with sensors in the thickness range 50-150 $\mu$m, with and without active edges,
 were investigated in view of the upgrade of the ATLAS pixel system for HL-LHC.
Very good interconnection yield between sensors and chips has been observed for these thin devices produced at CiS and ADVACAM.
Active edge sensors were investigated in terms of hit efficiency, both in the central active area and in the edge for different designs. 
The active area has been measured to extend up to the sensor physical perimeter for the design with only a floating guard ring  whereas the hit efficiency goes to zero in correspondence to the bias ring in the devices where this structure has been implemented.
An active edge sensor was irradiated at a fluence of $10^{15}$ \neqcm and increasing the bias voltage up to 250 V, the same tracking performance as in the not irradiated case could be obtained.
To allow for an investigation of the hit efficiencies at high pseudo-rapidity at 
HL-LHC for the 50x50 $\mu$m$^2$ pixel cell, FE-I4 modules were
tested in beam tests. The cluster properties were analyzed and a
good hit efficiency extracted for the single pixels, with the better performance observed for the thinner sensors, in the thickness range
100-200 $\mu$m.
\acknowledgments
This work has been partially performed in the framework of the CERN RD50 Collaboration. The
authors would like to thank A. Dierlamm for the irradiation at KIT, V. Cindro for the irradiations at JSI and
 L. Gonella for the irradiations at the MC40 Cyclotron of the University of Birmingham.
Supported by the H2020 project AIDA-2020, GA no. 654168.

% We suggest to always provide author, title and joucisrnal data:

\end{document}